# Height of Shock Formation in the Solar Corona Inferred from Observations of Type II Radio Bursts and Coronal Mass Ejections

N. Gopalswamy, H. Xie, P. Mäkelä, S. Yashiro, S. Akiyama
*Code 671, NASA Goddard Space Flight Center, Greenbelt, MD 20771, USA*
W. Uddin. A. K. Srivastava, and N. C. Joshi
*ARIES Nainital, Manora Peak, Nainital - 263 129, India*
R. Chandra
*Kumaun University, Nainital - 263 129, India*
P. K. Manoharan, K. Mahalakshmi, and V. C. Dwivedi
*TIFR/NCRA Radio Astronomy Center, P.O. Box 8, Ootacamund 643001, India*
R. Jain and A. K. Awasthi
*Physical Research Laboratory,* Navrangapura, *Ahmedabad - 380009, India*
N. V. Nitta and M. J. Aschwanden
*Lockheed Martin Solar and Astrophysical Laboratory, 3251 Hanover St, Palo Alto, CA 94304, USA*
D. P. Choudhary
*California State University, 18111 Nordhoff Street, Northridge, CA 91330, USA*

**Abstract:** Employing coronagraphic and EUV observations close to the solar surface made by the Solar Terrestrial Relations Observatory (STEREO) mission, we determined the heliocentric distance of coronal mass ejections (CMEs) at the starting time of associated metric type II bursts. We used the wave diameter and leading edge methods and measured the CME heights for a set of 32 metric type II bursts from solar cycle 24. We minimized the projection effects by making the measurements from a view that is roughly orthogonal to the direction of the ejection. We also chose image frames close to the onset times of the type II bursts, so no extrapolation was necessary. We found that the CMEs were located in the heliocentric distance range from 1.20 to 1.93 solar radii (Rs), with mean and median values of 1.43 and 1.38 Rs, respectively. We conclusively find that the shock formation can occur at heights substantially below 1.5 Rs. In a few cases, the CME height at type II onset was close to 2 Rs. In these cases, the starting frequency of the type II bursts was very low, in the range 25 – 40 MHz, which confirms that the shock can also form at larger heights. The starting frequencies of metric type II bursts have a weak correlation with the measured CME/shock heights and are consistent with the rapid decline of density with height in the inner corona.

## 1. Introduction

Type II solar radio bursts appear as slowly drifting features in radio dynamic spectra and are produced by electrons accelerated by MHD shocks (Nelson and Melrose, 1985). Although flare blast waves and CME-driven shocks have been considered as the source of type II bursts, recent observations indicate that all type II bursts may be due to CMEs (see Gopalswamy 2011 and references therein for a discussion on this topic). The CME height at the time of type II burst onset indicates the time the CME becomes super-Alfvénic to drive a fast mode MHD shock. In other words, the CME height at the time of type II onset indicates where and when the corresponding shock forms in the corona. The height of shock formation is important in understanding particle acceleration by shocks, especially the release time of solar energetic particles and their charge state properties (e.g. particles accelerated low in the corona may be subject to additional stripping – see Mewaldt et al. 2012).

The best set of CME observations became available from the Solar and Heliospheric Observatory (SOHO) mission around the beginning of solar cycle 23, so a detailed comparison between metric type II bursts and CMEs became possible. Gopalswamy et al. (2005) found that the average CME height at the time of the associated type II burst was ~2 Rs, which was obtained by back-extrapolating the CME height-time plot from measurements made beyond 2.5 Rs to the time of the type II burst. Unfortunately, in the height range between 1 and 2.5 Rs CMEs accelerate from rest and the speed changes rapidly compared to the constant speed measured in the outer field of view of SOHO's Large Angle and Spectrometric Coronagraph (LASCO) telescope C2. Thus the extrapolation based on LASCO data is subject to large errors. Case studies involving measurements closer to the Sun have suggested a lower CME height at type II onset. Cliver et al. (2004) used SOHO/LASCO/C1 data to show that the 1997 November 6 CME was at a height of about 1.3 Rs at the time of metric type II onset. The ground-based Mauna Loa Solar Observatory's K-Coronameter also observes the corona closer to the surface. Gopalswamy et al. (2012a) reported that the 2003 November 2 CME first appeared in the K-Coronameter field of view at a height of 1.67 Rs. The metric type II appeared ~5 minutes before the first appearance and hence the CME height is expected to be much lower than 1.67 Rs. MLSO height-time measurements were used to estimate the CME height as 1.23 Rs at type II burst onset. The Solar Terrestrial Relations Observatory (STEREO) mission's coronagraph COR1 has its inner edge of the field of view down to 1.4 Rs. Using COR1 observations Gopalswamy et al. (2009) found that the CME height at the time of type II burst onset was <1.5 Rs in many cases. Veronig et al. (2010) studied an EUV wave event observed by STEREO. From their height-time plot of the wave dome, the shock height can be estimated to be ~1.5 Rs (after correcting for projection effects). In another study, Gopalswamy et al. (2012a) determined the average initial acceleration of CMEs by combining the LASCO CME speed and the flare rise time to determine the CME height at type II onset to be in the range 1.38 to 1.53 Rs. Interferometric imaging of type II bursts at 109 MHz has shown that the burst centroids are typically in this height range (Ramesh et al. 2012). In a recent case study, Gopalswamy et al. (2012b) showed that the 2010 June 13 type II burst occurred precisely when the shock appeared in the EUV images obtained by the Atmospheric Imaging Assembly (AIA) on board the Solar Dynamics Observatory (SDO). In this event, the CME flux rope and the shock were at heliocentric distances of 1.17 and 1.19 Rs, respectively, even smaller than the heights obtained from COR1 data (Gopalswamy et al. 2009) and from the flare acceleration method (Gopalswamy et al.

2012a). The 2010 June 13 CME eruption was a limb event, so the shock height at type II onset was obtained directly by measuring the leading edge of the shock in EUV images. Occurrence of the metric type II burst and the EUV shock have been found simultaneous (Gopalswamy e al. 2012). There has been some controversy regarding the association between EUV waves and shocks (see e.g., Gallagher and Long, 2011). In our opinion, the controversy arises because of incomplete consideration given to the existence of CMEs in the EUV wave events. With better observations from STEREO/EUVI and SDO/AIA, as noted above in the case of the 2010 June 13 event, it has become clear that the EUV wave surrounds the associated CME and hence is driven by the CME. Taking that the EUV wave is a shock when it is associated with a type II burst, one can get the shock height at type II burst onset assuming that the shock has a hemispherical shape surrounding the CME. This means the shock height is simply the radius of the projected circle of the EUV wave when it occurs on the disk. We call this the "wave diameter method", more suitable for disk events as opposed to the leading edge method, which is more suitable for limb events. Note that this method is applicable during the initial stage of the CME evolution when the lateral and radial expansions are similar. Later on, the lateral expansion may cease leading to smaller lateral speed (Veronig et al. 2010). The type II burst just provides the time marker as to when the shock forms. The type II burst can occur anywhere on the shock front. The CME height and shock height at the nose are the primary parameters we are interested in. In the wave diameter method, the outermost disturbance is the shock, because it is assumed to be a bow wave surrounding the CME. Simultaneous observations of the same event in limb and disk views support the wave diameter method. In this paper, we consider 32 type II bursts for which STEREO's Extreme Ultraviolet Imager (EUVI) and/or COR1 observations are available at the time of type II bursts, so the shock height can be determined without projection effects. We primarily use EUVI images obtained at 195 Å. In the following we refer to instruments on STEREO ahead (STA) and behind (STB) with suffixes A and B, respectively.

Section 2 describes the wave diameter and leading edge methods and lists the 32 events considered for the study. Section 3 presents results of the analyses: distributions of CME heights at metric type II onset and the starting frequencies of type II bursts including the relation between the two quantities. Section 4 discusses the results and the study is summarized in section 5.

## 2. Data Selection and Method

We considered metric type II bursts from cycle 24 when the STEREO spacecraft started observing in quadrature with spacecraft along the Sun-Earth line. More than 100 type II bursts were reported in Solar Geophysical Data (ftp://ftp.ngdc.noaa.gov/STP/SOLAR_DATA/SOLAR_RADIO/SPECTRAL/) and the recent ones from the NOAA Space Weather Prediction Center: http://www.swpc.noaa.gov/ftpmenu/warehouse.html. The primary criterion used in selecting the type II bursts for this study is that the solar source must be such that the associated eruption is fully observed as a disk or limb event in EUVI or COR1 Images. Several limb events in STEREO view were observed as disk events by SDO/AIA, which we use for confirmation of the STEREO measurements.

### 2.1 Leading Edge Method

Figure 1 shows a EUVI frame taken on 2011 September 06 at 01:45:52 UT, close to the appearance of the type II burst at 01:46 UT in the Culgoora radio dynamic spectrum. The eruption occurred from close to the Earth-facing disk center (N14W07) and was accompanied by a M5.3 flare and a halo CME (speed ~782 km/s appearing at 02:24 UT – see cdaw.gsfc.nasa.gov/CME_list/halo/halo.html). The type II burst ended before the CME appeared in the LASCO field of view, so we need to resort to inner coronal imagers to observe the CME closer to type II onset. In EUV images, CMEs often appear similar to their white-light counterparts (Dere et al., 1997; Gopalswamy et al., 2012b). The nearest EUVI image from STA is shown in Figure 1(b). The CME leading edge was at a height of 1.29 Rs at the time of the type II burst. STA was ahead of Earth at W102 at the time of this event, so the eruption corresponds to E95 from STA view. This is right at the limb, so the measured height of the CME is devoid of projection effects. A fuzzy feature can be seen surrounding the CME, which is at a distance of 0.06 Rs from the CME. This feature is likely to be the shock sheath. Since the shock feature is not clear, we use the CME height as the main parameter indicating the height at which the type II burst forms.

## 2.2 Wave Diameter Method

The wave diameter method is illustrated in Figure 2 for the type II burst of 2011 January 28 at 01:01 UT. The EUVI A image taken at 01:00:41 UT shows the approximately circular EUV disturbance surrounding the eruption region close to the disk center in STA view. In Earth view, the eruption occurred at the west limb (N16W91) from NOAA active region 11149. STA was at W95, so the eruption appears as a disk event (W04) in STA view. We fit a circle to the outermost part of the disturbance and take its radius as the shock height above the solar surface (the assumption is that the disturbance expands spherically above the solar surface). In this case, the radius of the circle is 0.3 Rs, so the heliocentric distance of the shock becomes 1.3 Rs. The CME was observed as a west limb event by SDO/AIA and the height of the CME at 01:00:43 UT was 1.38 Rs, which agrees well with the height obtained by the wave diameter method.

Table 1 lists the 32 metric type II bursts from 2010 June 12 to 2012 April 24 that were relatively well observed by the STEREO instruments. Column 2 gives the date (year, month, day) and time of the metric type II burst. Column 3 gives the heliographic coordinates of the source region of the CME (Earth view) that resulted in the type II burst. Columns 4-6 give the name of the STEREO instrument used, the source longitude in STEREO view, and the time of the STEREO image in which the CME heights were measured. Most of the measurements were made in EUVI images (18 from EUVI B and 11 from EUVI A) and three from the COR1. Column 7 gives the measured height of the CME-driven shock (measured as the heliocentric distance). The leading edge method was used for seven events (identified in column 1 of Table 1); the wave diameter method was used for the remaining events. Note that the heights in this column correspond to the shock height when the wave diameter method was used. As noted in section 2.1, some of the values from the leading edge method may be just the height of the flux rope because the shock is not distinctly seen in the images. When the shock structure is distinctly observed, the standoff distance (the thickness of the shock sheath measured as the difference between the leading edges of the shock feature and the flux rope) becomes the error in the CME/shock height. For example, the in the case of the 2010 June 13 event, the standoff distance was ~0.02 Rs (Gopalswamy et al. 2012b), which is only 1.7% of the flux rope height. However, the difference is rather small, so we do not distinguish between shock and the driving CME. The last column in Table 1 lists the starting frequency of the fundamental component of the type II burst. In most cases, the fundamental-harmonic structure is present, so it is straightforward to get the starting frequency. In a few cases, there was no fundamental-harmonic structure. For limb events, we assumed the emission to be in the harmonic mode (the fundamental is preferentially absorbed by the foreground plasma). For disk events without fundamental-harmonic structure, we assumed the emission to be the fundamental. Fortunately, there were only a few such ambiguous events, as noted in Table 1.

**3. CME Height at Type II Burst Onset**

The CME heights range from 1.20 Rs to 1.93 Rs (see column 7 of Table 1). The distribution of CME heights is given in Fig. 3(a). The distribution is roughly symmetric with mean and median values of 1.43 and 1.38 Rs, respectively, except for three outliers that had a CME height of 1.93 Rs. The shock formation heights should also be indicated by the starting frequencies of type II bursts, the distribution of which is shown in Fig. 3(b). The mean and median starting frequencies are 102 and 85 MHz, respectively. Figure 3(c) shows that there is a weak correlation between the starting frequency (f) and the CME height (r). The correlation coefficient (CC) is 0.56 when the two high starting frequency events are excluded (inclusion of these two data points reduces CC to 0.53). The relationship can be fit to a power law: $f(r) = 307.87r^{-3.78} - 0.14$. Since f is the same as the local plasma frequency (due to plasma emission mechanism for type II bursts) this relation implies that the plasma density varies as $r^{-7.56}$

. Such a drop would be consistent with the large power law index of ~6 derived from eclipse observations (see e.g., Allen, 1947; Newkirk, 1967) and the universal frequency - drift rate relationship (Gopalswamy et al. 2009; Gopalswamy, 2011). There is a large scatter because the data points correspond to different days over a 3-year period in cycle 24 and the coronal densities are expected to have different distributions on different days. In fact, one can consider exponential fall off of the coronal electron density with a hydrostatic scale height. Converting the density to plasma frequency, we get the two curves in Fig. 3(d) for two coronal base densities ($n_{base}$) that differ by a factor of 10 and an electron temperature ($T_e$) of 2 MK. Most of the data points are encompassed by the two curves. Thus the measurements of shock formation height are consistent with the expected decline of the plasma frequency with height in the corona.

The two high starting frequency events have CME heights within 1.34 Rs, which may indicate propagation of the shock through high-density magnetic structures (see e.g., Pohjolainen et al. 2008). For the three events with the largest CME heights at type II burst onset, the measurements were made from COR1 images using the leading edge method. The three events correspond to the last three data points in Fig. 3(c) and clearly correspond to low starting frequencies. The measurement time is within 28 seconds of the type II onset for the 2011 March 7 event. For the other two events, the measurement times are 2 (2011 August 2) and 3 min (2011 September 30) after the onset of the type II burst. We examined the three events in more detail to understand the large deviations. The 2011 March 07 eruption was a limb event for both STA (E111) and STB (W80) so the leading edge method should give an accurate height of the CME at type II onset. As shown in Fig. 4, the fundamental (F) component of the type II burst starts at 25 MHz. The dynamic spectrum from the Green Bank Radio Spectrometer clearly shows the fundamental harmonic structure and no emission at higher frequencies. At much higher frequencies (above 100 MHz, not shown) there was type IV emission, but no type II burst. Thus we conclude that the shock formed at a larger height in this case.

The 2011 August 2 type II burst also started at low frequencies: the fundamental emission starts at 40 MHz with the harmonic component at 80 MHz in the Culgoora radio dynamic spectrum. Finally, the 2011 September 30 event with larger CME height at type II onset is also a low starting frequency case. In the Green Bank Radio Burst Spectrometer data, the fundamental starts at 30 MHz and drifts down to about 10 MHz. The harmonic component is also clearly seen and starts at 60 MHz. The starting time of the burst is also at 19:09 UT, rather than the 19:08 UT reported in the Solar Geophysical Data. The CME had already appeared in the STB COR1 field of view at 19:05:56 at which time there was no type II burst. Therefore, the large CME height at type II onset can be safely attributed to the fact that the shock forms at a larger height in all the three cases.

### 3.1 Comparison with SDO Measurements

We already noted that the leading edge method applied to the SDO/AIA 193 Å images for the 2011 January 28 at 01:01 event confirmed the EUVI measurement for the same event using the wave diameter method (Fig. 2). The agreement was also good for the 2010 June 13 event (Gopalswamy et al. 2012b). We were also able to apply the wave diameter method to six events observed by SDO/AIA as disk events. The CME heights obtained using SDO and EUVI data are compared in Table 2. We see that there is very good agreement between the wave diameter (SDO) and leading edge (EUVI) measurements. The difference in CME heights between the two measurements is in the range -2% to +6.7%.

### 4. Discussion

The range of CME heights at type II burst onset in the present study is 1.20 Rs to 1.93 Rs with a mean value of 1.43 Rs. This range is consistent with the starting frequencies of type II bursts. The range and mean values of the CME height are remarkably similar to those obtained from the acceleration method in Gopalswamy et al. (2012a) for 16 ground level enhancement events of solar cycle 23. In a recent study, Ramesh et al. (2012) reported the heights of the type II burst centroids measured directly from two-dimensional radio images obtained at 109 MHz using the Gauribidanur radioheliograph. Out of the 41 bursts they reported, 7 had originated from close to the limb (central meridian distance 80 to 90 degrees). The type II heights ranged from 1.2 to 1.9 Rs for the seven events with an average value of 1.44 Rs. The onset of metric type II bursts when the CME reaches the height range 1.2 – 1.9 Rs is consistent with the minimum in Alfven speed profile around 1.5 Rs caused by the rising profile due to the quiet corona (Mann et al., 1999) and the declining profile due to the active region (Gopalswamy et al. 2001). The present study is a good demonstration of the robustness of the Alfven speed profile reported in Gopalswamy et al. (2001), Mann et al. (2003), and Warmuth and Mann (2005). It must be pointed out that the bursts considered in the paper belong to the rise to maximum phase of cycle 24. It is possible that some of the very low shock formation heights reflect the fact that the plasma levels are closer to the surface than usual (Gopalswamy et al., 2009). A systematic study is needed by selecting bursts with similar starting frequencies from the same of phase of different cycles.

It must be noted that the EUVI or COR1 images used were obtained within 3 minutes of the metric type burst onset (EUVI images have a cadence of 5 min), generally after the burst onset. The minimum and maximum deviations are 1s and 176s, respectively, with an average value of 89s. This means a slight overestimate of the CME height. Making use of the fact that the typical CME speed in the acceleration phase around 1.5 Rs is ~500 km/s, we estimate that the CME would have traveled a distance of ~0.06 Rs in 89s. Compared to the mean distance of 1.43 Rs, this deviation represents an error of 4.5%. The maximum deviation is 176s, which corresponds to a height difference of 0.12 Rs, which is a 8.8% deviation from 1.43 Rs.

Throughout the paper, we have taken the CME height to be the same as the shock height and the type II burst height. This is strictly not true because the shock is expected to be located ahead of the CME and the type II burst is located at the shock front. We can estimate the expected error from the case study reported in Gopalswamy et al. (2012b) on the 2010 June 13 CME and shock. The standoff distance (0.03 Rs) measured directly in EUV images was a small fraction of the CME height of (1.17 Rs). The shock height in this case is an underestimate by 2.6%. We do not expect the error to be much larger in the small number of cases for which the leading edge method was used. For the cases where the wave diameter method was used, the measurement corresponds to the shock, so there is no underestimate. In Table 2, we showed that for several events we were able to use both leading edge and wave diameter methods. The maximum discrepancy is 6.7%.

Finally, we note that the type II radio emission can originate anywhere on the shock front: at the nose or flanks depending on which location is favorable for electron acceleration. In a recent study, Ramesh et al. (2012) computed the position angle difference between CME nose and type II burst centroid for a set of 41 events. The type II bursts were imaged by the Gauribidanur Radioheliograph at a frequency of 109 MHz, which is very close to the mean starting frequency of type II bursts in our sample (see Fig. 3b). The associated eruptions occurred close to the limb, so the measured heights and position angles are not affected by projection effects. They found that the position angles of type II burst centroid deviated from the central position angle of the associated CME by 0 to 46 degrees with an average value of 14 degrees. For an average CME height of 1.43 Rs (see Fig. 3a), the average position angle offset of the type II burst corresponds to a height difference of only 0.04 Rs, or a 3% difference. Thus our assumption that the CME/shock height corresponds to the height of the type II burst is reasonable.

**5. Summary**

We determined the CME/shock height at the time of metric Type II burst onset using the wave-diameter and leading-edge methods for 32 events during the interval 2010-2012. We made use of the STEREO EUV and coronagraphic observations, which enabled us determine the CME height without projection effects and significant extrapolation. We primarily used STEREO/EUVI images and in three cases, COR1 images. We used the leading-edge method or the wave-diameter method depending on the location of the source region on the Sun. The coronal images were chosen as close to the type II burst onset as possible to avoid extrapolation. We found that the shock nose is located in the heliocentric distance range from 1.20 to 1.93 Rs, with mean and median values of 1.43 and 1.38 Rs, respectively. This result confirms and refines an earlier result obtained by Gopalswamy et al. (2009) who investigated several type II burst events that occurred during the early phase of the STEREO mission and found the shock height to be <1.5 Rs. Shock formation close to the surface has important implications for theories of particle acceleration at the Sun. The results are also consistent with the shock distances obtained from another method that makes use of the CME initial acceleration during ground level enhancement in solar energetic particle events (Gopalswamy et al., 2012b). The CME/shock heights are consistent with the distribution of type II starting frequencies in the fundamental mode. The scatter plot between the starting frequencies and the CME/shock heights mimics the distribution of plasma frequencies with height. The power law dependence of the plasma frequency with height is consistent with the rapid fall off of density in the inner corona.

**Acknowledgments.** This study was conducted as a part of the Indo-US Science and Technology Forum's Joint Center on Solar Eruptive Events. We acknowledge the use of radio dynamic spectra made available on line at Culgoora, Hiraiso, Green Bank, and the RSTN data from NGDC. This work was supported by NASA's LWS TRT program. PKM was partly supported by ISRO's CAWSES-India Program.

Tables

Table 1. List of metric type II bursts for which the height of formation was determined from STEREO data

| S. No. | Type II Burst Date UT | Source Location | STEREO Instrument | ST Source Longitude | STEREO Image UT | Shock Height (Rs) | Starting f (MHz) |
|---|---|---|---|---|---|---|---|
| 1 | 20100612 00:56 | N22W57 | EUVI A | E16 | 00:55:59 | 1.20 | 105 |
| 2 | 20100613 05:37 | S25W84 | EUVI A | W10 | 05:38:29 | 1.21 | 155 |
| 3 | 20100807 18:08 | N12E31 | EUVI B | W40 | 18:10:56 | 1.45 | 90 |
| 4 | 20101016 19:14 | S19W29 | EUVI A | E54 | 19:15:45 | 1.36 | 90[b] |
| 5 | 20101103 06:13 | S00E90 | EUVI B | E08 | 06:15:42 | 1.67 | 48[c] |
| 6 | 20101103 12:14 | S00E90 | EUVI B | E08 | 12:15:42 | 1.34 | 280 |
| 7[a] | 20101112 01:37 | S22W08 | EUVI A | E76 | 01:38:11 | 1.56 | 100 |
| 8 | 20101215 14:38 | N31W60 | EUVI A | E25 | 14:40:39 | 1.41 | 60 |
| 9 | 20101231 04:23 | N17W54 | EUVI A | E31 | 04:25:39 | 1.20 | 170 |
| 10 | 20110127 12:08 | N17W91 | EUVI A | W04 | 12:10:41 | 1.49 | 40[b] |
| 11 | 20110128 01:01 | N16W91 | EUVI A | W04 | 01:00:52 | 1.30 | 70 |
| 12 | 20110211 21:46 | N00E90 | EUVI B | W03 | 21:45:42 | 1.26 | 200[c] |
| 13[a] | 20110213 17:35 | S20E04 | EUVI B | W89 | 17:35:43 | 1.22 | 400 |
| 14[a] | 20110214 13:03 | S20W12 | EUVI A | E74 | 13:05:43 | 1.47 | 42[c] |
| 15 | 20110215 01:52 | S21W21 | EUVI A | E66 | 01:50:44 | 1.21 | 150 |
| 16 | 20110216 14:23 | S21W37 | EUVI A | E49 | 14:25:44 | 1.34 | 175[c] |
| 17[a] | 20110307 14:25 | N11E14 | COR1 B | W80 | 14:25:16 | 1.93 | 25 |
| 18 | 20110308 03:44 | S00E90 | EUVI B | W04 | 03:45:47 | 1.46 | 80[c] |
| 19 | 20110325 23:14 | S15E45 | EUVI B | W50 | 23:15:54 | 1.38 | 125 |
| 20 | 20110511 02:27 | N00W45 | EUVI A | E47 | 02:30:57 | 1.44 | 75 |
| 21 | 20110530 11:06 | S19E57 | EUVI B | W36 | 11:05:58 | 1.39 | 80[d] |
| 22[a] | 20110802 06:08 | N16W20 | COR1 A | E80 | 06:10:26 | 1.93 | 45 |
| 23 | 20110810 16:07 | N25E70 | EUVI B | W23 | 16:05:56 | 1.32 | 30 |
| 24 | 20110828 04:20 | N00W90 | EUVI A | E12 | 04:20:53 | 1.38 | 90 |
| 25[a] | 20110906 01:46 | N14W07 | EUVI A | E95 | 01:46:13 | 1.29 | 150 |
| 26[a] | 20110930 19:08 | N13E02 | COR1 B | W95 | 19:10:17 | 1.93 | 32 |
| 27 | 20111119 01:29 | N00W90 | EUVI A | E16 | 01:30:42 | 1.45 | 90 |
| 28 | 20120105 07:11 | N10W71 | EUVI A | E36 | 07:10:40 | 1.33 | 85[d] |
| 29 | 20120118 23:23 | N25W44 | EUVI A | E63 | 23:25:41 | 1.53 | 45 |
| 30 | 20120120 21:16 | N25W71 | EUVI A | E36 | 21:15:41 | 1.34 | 50 |
| 31 | 20120324 08:45 | S26E81 | EUVI B | W37 | 08:46:31 | 1.49 | 50 |
| 32 | 20120424 07:48 | N14E68 | EUVI B | W50 | 07:45:54 | 1.51 | 45 |

[a]Height measurements from leading-edge method; [b]May be slightly higher; [c]assuming harmonic observed; [d]assuming fundamental observed

Table 2. List of events with SDO measurements

| S. No | Date & Time | SDO Time | CME Height Rs (SDO) | CME Height Rs (EUVI) | Deviation (%) |
|---|---|---|---|---|---|
| 7 | 20101112 01:37 | 01:39:07 | 1.48 | 1.56 | 5.1 |
| 13 | 20110213 17:35 | 17:35:07 | 1.21 | 1.22 | 0.8 |
| 14 | 20110214 13:03 | 13:04:19 | 1.50 | 1.47 | -2.0 |
| 17 | 20110307 14:25 | 14:23:55 | 1.81 | 1.93 | 6.2 |
| 22 | 20110802 06:08 | 06:08:19 | 1.80 | 1.93 | 6.7 |

**Figures**

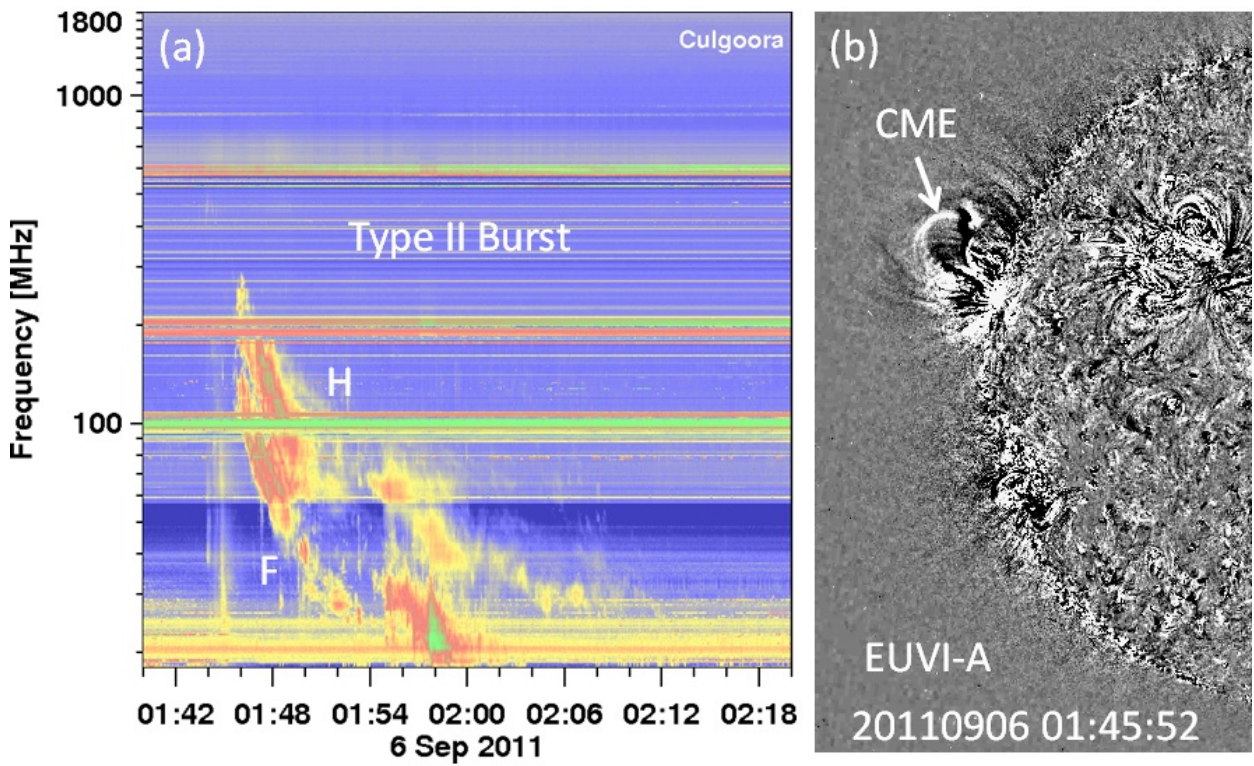

Figure 1. (a) Dynamic spectrum from the Culgoora radio observatory showing the type II burst with fundamental (F) and harmonic (H) structure. The fundamental component starts around 150 MHz. (b) A section of the nearest EUVI A image showing the CME. The CME height can be directly measured from this frame as 1.29 Rs.

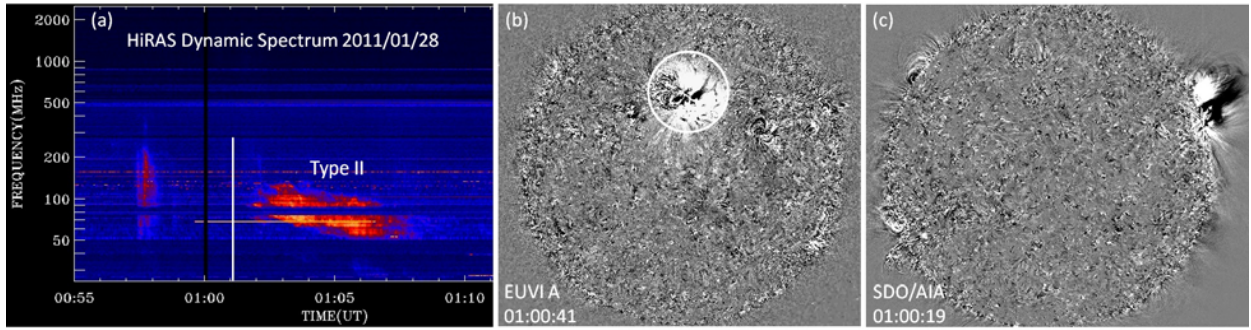

Figure 2. (a) Type II burst from the Hiraiso Radio Spectrograph (Japan) showing the type II burst, which is prominent in the harmonic component. The fundamental component is weak and can be seen at the bottom of the spectrum. The type II burst starts around 01:01 UT. (b) EUVI A showing the EUV wave at 01:00:41 UT, which is very close to the onset of the type II burst. A circle can be fit to the outermost part of the EUV wave, which is the base of the hemispherical shock that surrounds the CME. (c) SDO/AIA image at 01:00:19 UT showing the CME above the northwest limb. The CME partially left the SDO/AIA field of view, but the height at the nose can be measured as 1.38 Rs.

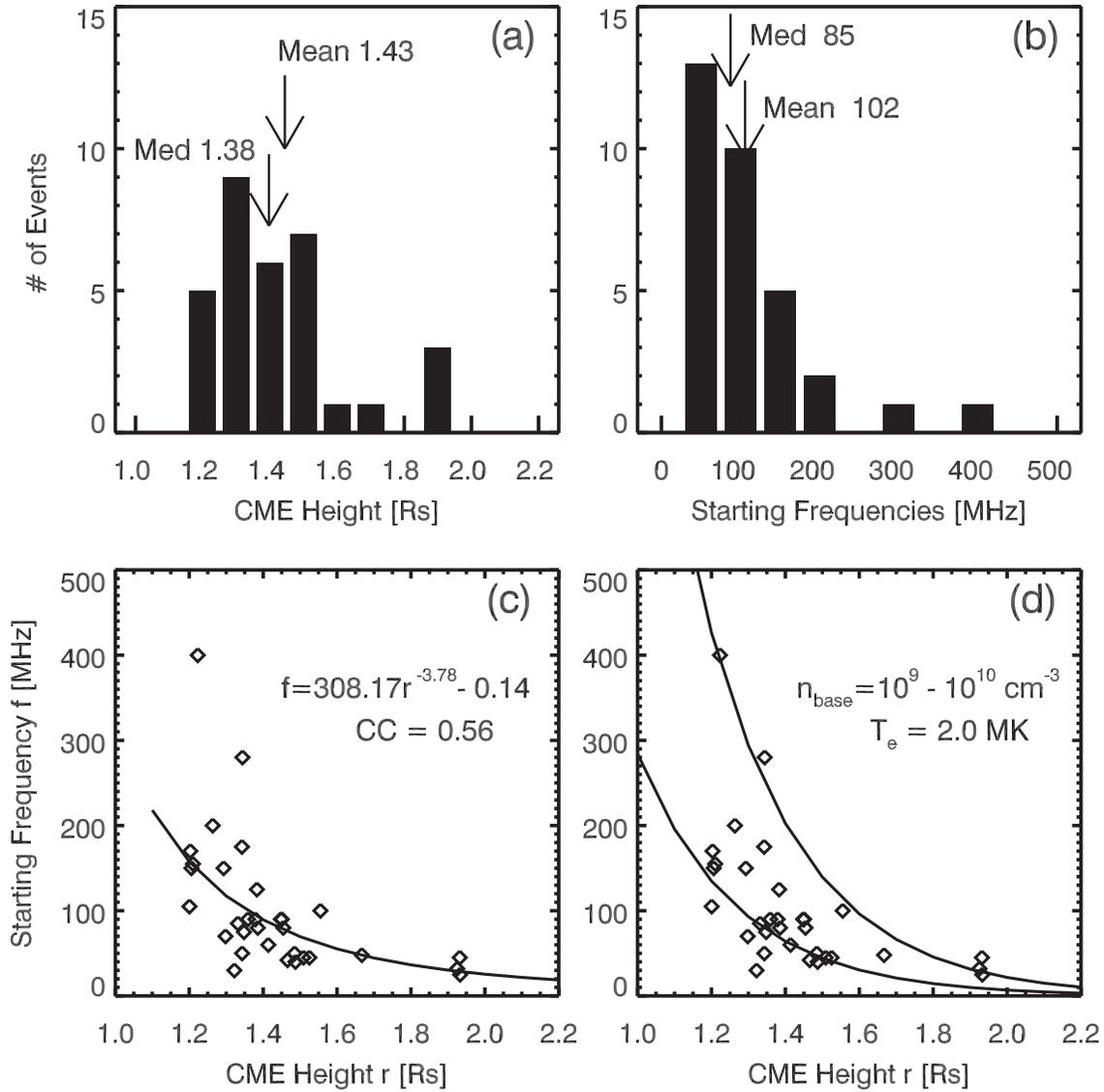

Figure 3. (a) Distribution of CME heights at the time of type II burst onset. (b) Distribution of starting frequencies of type II bursts in the fundamental mode. The mean and median values are noted on the plot. The three outliers in the height distribution are all obtained using the leading edge method applied to STEREO/COR1 observations. (c) Scatter plot between CME height and the Type II starting frequency along with a power law fit to the data points. The correlation coefficient (CC) = 0.56 was obtained excluding the two high starting frequency cases. Inclusion of these two events reduces CC slightly to 0.53. (d) Plasma frequency vs. height plotted for two coronal base densities: $n_{base} = 10^9$ cm$^{-3}$ (left) and $10^{10}$ cm$^{-3}$ (right) curves, assuming an electron temperature ($T_e$) of 2 MK.

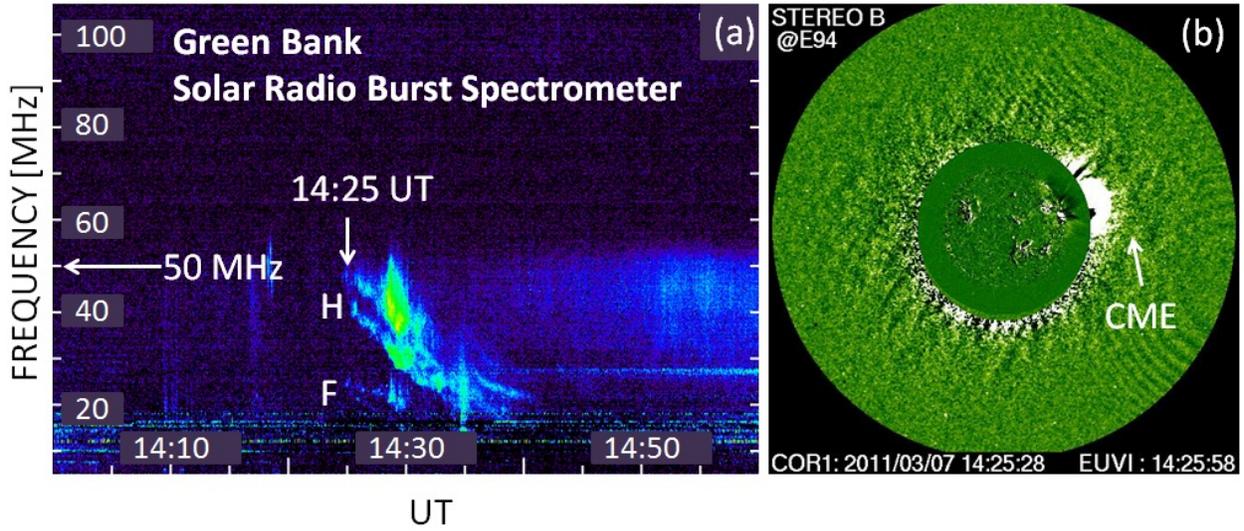

Figure 4. (a) Dynamic spectrum from the Green Bank Radio Spectrometer showing the type II burst starting around 14:25 UT with fundamental (F) – harmonic (H) structure. (b) COR1 B image at 14:25 UT showing the CME above the northwest limb in STB view. The COR1 image is superposed on EUVI 195 Å difference image, which shows the eruption as a dimming feature. STB was located at E94.